\documentclass[USenglish,twocolumn]{article}
\usepackage[margin=2cm]{geometry}
\usepackage{authblk}
\usepackage{booktabs}
\usepackage{array}
\usepackage{anyfontsize}
\usepackage{amsmath}  
\usepackage{dsfont}   
\usepackage{amsmath,amssymb}
\usepackage{graphicx}
\usepackage{xcolor}
\usepackage{epstopdf}
\usepackage{hyperref}
\usepackage[normalem]{ulem}

\newcommand*\add[1]{#1}
\newcommand*\cut[1]{}

\providecommand{\keywords}[1]{\textbf{Keywords:} #1}

\begin{document}

\author[1,*]{Egor Manuylovich}
\author[1]{Dmitrii Stoliarov}
\author[2]{David Saad}
\author[1]{Sergei K. Turitsyn}

\affil[1]{Aston Institute of Photonic Technologies (AiPT), Aston University, UK}
\affil[2]{Aston Centre for Artificial Intelligence Research and Application, Aston University, UK}
\affil[*]{e.manuylovich@aston.ac.uk}

\title{Optical neuromorphic computing \add{via temporal up-sampling and trainable encoding on a}\cut{using} telecom device platform\cut{ for nonlinear mapping to high-dimensional feature space}}

\date{}

\maketitle

\abstract{
Mapping input signals to a high-dimensional space is a critical concept in various neuromorphic computing paradigms, including models such as Reservoir Computing (RC) and Extreme Learning Machines (ELM). We propose using commercially available telecom devices and technologies developed for high-speed optical data transmission to implement these models  through nonlinear mapping of optical signals into a high-dimensional space where linear processing can be applied. We manipulate the output feature dimension by applying temporal up-sampling (at the speed of commercially available telecom devices) of input signals and a well-established wave-division-multiplexing (WDM). Our up-sampling approach utilizes a  trainable encoding mask, where each input symbol is replaced with a structured sequence of masked symbols, effectively increasing the representational capacity of the feature space. This gives remarkable flexibility in the {\it dynamical phase masking} of the input signal. We demonstrate  this approach in the context of RC and ELM, employing readily available photonic devices, including a semiconductor optical amplifier and nonlinear Mach-Zender interferometer (MZI). We investigate how nonlinear mapping provided by these devices can be characterized in terms of the increased controlled separability and predictability of the output state.
}
\\
\\


\section{Introduction} 
There is a growing interest in unconventional approaches to computing, as traditional digital computing is reaching its fundamental limitations~\cite{keyes1981fundamental, jaeger2021towards}, particularly due to the unsustainable power consumption of machine learning approaches~\cite{Strubell_Ganesh_McCallum_2020}.
Reservoir computing (RC)~\cite{RC01,RC02} and extreme learning machine (ELM)~\cite{ELM01} are two popular unconventional (non-digital) computing concepts implicitly based on nonlinear mapping of the input into a high dimensional output where it can be processed using simple and efficient linear algorithms. For practical implementation purposes, operating only with a linear readout layer is a substantial advantage, due to the straightforward processing and low computational complexity required. The main difference between the two methods is that the RC architectures exploit recurrent connections, creating memory in the system, while ELM is a feed-forward approach that does not use any memory. 

The idea of mapping input signals into a high-dimensional output for processing is rooted in the foundations of information theory, but also features in machine learning approaches, such as, support vector machines (SVM)~\cite{cristianini2000introduction} (with predetermined nonlinear mapping) and reservoir computing, which uses a recurrent neural network as an uncontrolled nonlinear mapping. Difficult computing tasks are made easier by transforming them nonlinearly to a higher dimensional space where linear processing can be applied. In this approach computing is treated as structuring the nonlinear mapping, instead of processing of structures as in traditional computing. 
Note that creating high-dimensional space available for the mapping is not sufficient in itself, since the output signal should be effectively spread across this feature space, without making many dimensions redundant.
Therefore, the important challenge is to ensure a high effective dimensionality of the output, that is
quantified by a set of linearly independent variables.

Manipulating the effective dimensionality of the feature space by a nonlinear transformation of the streamed temporal signal paves the way for a range of non-conventional computing methods. In particular, it is well suited for manipulating temporal continuous (analog) signals that are naturally generated in sensing, imaging and a number of other applications. Analog information processing is one of the key pillars of unconventional computing. Analog computing which has a long history, is experiencing a resurgence due to its superior power efficiency and capability of parallel processing \cite{analog01}. Analog computing is well suited for tasks that require continuous inputs and outputs. The effective dimensionality can be controlled in different ways (e.g. temporal sampling or frequency filtering) projecting the infinite-dimensional space of the analog signal after nonlinear transformation onto a well-separated set of features.  Additionally, effective dimensionality can be controlled by the parameters and characteristics of the nonlinear transformation.

Analog information processing can be implemented in a variety of physical systems trained to learn complex features.
In more general terms, natural, human-engineered physical, chemical, and biological systems can be used as substrates to realize computing algorithms
(see, e.g. \cite{Phys00,Phys01,Phys02,Phys03,Phys04} and references therein). 

Ultra-fast nonlinear photonic systems, in particular, are attractive for the implementation of unconventional computing approaches due to their relatively low power dissipation and capability of parallel signal processing. Recently, there has been a great deal of interest in the development of photonic-based ELMs and RCs (see, e.g., \cite{pierangeli2021photonic, PELM02, PELM03, Aston01, PELM04, WDM01sorokina,WDM01wang, lupo2021photonic, RCOPT03, biasi2023array} and references therein). Several notable works have demonstrated the potential of photonic reservoir computing in leveraging existing telecom technologies. A 16-node square mesh reservoir on a silicon photonics chip was implemented, capable of executing Boolean operations and header recognition tasks \cite{vandoorne2014experimental}. A parallel photonic reservoir computing approach using semiconductor optical amplifiers was also investigated, demonstrating competitive performance in speech recognition tasks  \cite{vandoorne2011parallel}. A unified framework for reservoir computing and extreme learning machines was developed using a single nonlinear neuron with delayed feedback, realizable in optoelectronic and all-optical implementations \cite{ortin2015unified}. Another study experimentally implemented reservoir computing with a nonlinear optoelectronic oscillator, achieving high performance on spoken digit recognition and time series prediction tasks \cite{larger2012photonic}. A high-speed photonic reservoir computing system based on InGaAsP microring resonators was demonstrated for efficient all-optical pattern recognition in dispersive Fourier imaging \cite{mesaritakis2015high}.  

Light possesses a rich set of degrees of freedom that can carry information. In optical communications, parameters such as amplitude, phase, wavelength, and polarization are routinely used to encode signals, with spatial division multiplexing emerging as an important technique for increasing data transmission rates further. Several examples of optical neuromorphic computing exploiting spectral signal multiplexing have already been explored, including implementations based on optical reservoir computing\cite{WDM01sorokina, WDM02, WDM05} as well as other optical neuromorphic architectures \cite{WDM01wang, WDM001, WDM03, WDM04}. Recent works demonstrated that WDM enhances photonic reservoir computing by enabling parallel processing and increasing computational capacity. A Fabry–Perot semiconductor laser-based RC leveraged multiple wavelength channels to improve signal equalization in optical communications \cite{li2023scalable}. A microring resonator-based RC exploited WDM for simultaneous multi-task processing, showing its potential for efficient parallel computing \cite{giron2024multi}. A waveguide-based RC demonstrated improved nonlinear signal equalization across multiple WDM channels \cite{gooskens2022wavelength}. Additionally, WDM has been shown to enhance the memory capacity of RC without requiring external optical feedback by using wavelength-multiplexed delayed inputs in microring-based architectures \cite{castro2024wavelength}. In this work, we show that WDM can also be used to enable a faster encoding mask, surpassing the speed limitations of single-channel implementations.

In the temporal domain, up-sampling was used in \cite{takano2018compact}. However, the advantages of a faster output sampling rate were not leveraged because of the low bandwidth of the readout system. In this work, we demonstrate the advantage of up-sampling in enhancing the performance metrics on different tasks. Also, we would like to point out that unlike the previously studied delay-based  RC implementations, \cite{takano2018compact, nakajima2021scalable}, we propose approach without additional optical delay, while providing comparable processing capacity. 
Note that in the majority of the demonstrated optical reservoir computing and ELMs, only one of these degrees of freedom was used. 

Photonic ELMs are characterized by their feed-forward architecture, which features untrained internal connections and trained output weights. In the optical domain, the non-trainable, nonlinear signal transformation characteristic of ELM  is performed automatically and cost-effectively through physical signal propagation and registration, leveraging the inherent properties of light and nonlinear optical components and systems. Photonic ELMs have been demonstrated in various setups, including a free-space optical propagation scheme \cite{pierangeli2021photonic} and frequency-multiplexed fiber framework \cite{lupo2021photonic}. The implementation using the array of microresonators on an integrated silicon chip \cite{biasi2023array} achieved notable success in both binary and analog tasks, underscoring the potential of ELMs in photonics for efficient and high-performance machine learning applications.  The efficacy of photonic ELMs has been further enhanced by employing feedback alignment for training the input mapping, see, e.g., \cite{kilic2021training}.
 
Recent applications of optical reservoir computing include: modulation format identification in fiber communications using single dynamical node-based photonic RC \cite{cai2021modulation}, machine learning based on RC with time-delayed optoelectronic and photonic systems \cite{chembo2020machine}, photonic neuromorphic technologies in channel equalization \cite{argyris2022photonic}, analog optical computing for artificial intelligence \cite{wu2022analog} and many others.  

In this work, we propose and demonstrate how devices and technologies developed for optical data transmission can be used for computing applications. \add{While individual techniques like high-dimensional mapping, input masking, WDM, and specific nonlinear elements have been explored previously, our contribution focuses on a novel combination of these elements within a flexible framework designed for effective nonlinear mapping.} The \add{combined use of} signal \cut{up-sampling in the temporal domain}\add{feature-space expansion via temporal up-sampling} and WDM technology\add{ for parallel encoding} gives a great degree of flexibility in designing the structure of a high-dimensional output. \add{A key element of our approach is the incorporation of a trainable input encoding mask. This trainable mask provides a crucial advantage, allowing for task-specific optimization of the input signal representation before it undergoes nonlinear transformation, thereby enhancing the overall representational capacity and performance of the system.} Multiple spectral channels can range from coarse WDM (systems with fewer than eight active wavelengths per fiber) to dense WDM (DWDM). DWDM can offer standard telecom solutions with a number of channels varying from tens to hundreds with typical (but also variable) channel spacing of 50GHz or 100GHz within the so-called optical fiber C-band (spectral interval from 1530 nm to 1565 nm).
Traditional DWDM systems exploit wavelength-selective switches designed with fixed 50GHz or 100GHz filters. Using other fiber spectral bands, DWDM can be extended to thousands of channels. In the temporal domain, data streams with symbol rates as high as 32, 64 Gbaud (and more) in a single fiber can be produced with standard components. Thus, commercially available telecom devices can be utilized to produce a huge dimensional output feature space using only standard conventional technology. Numerous non-linear optical elements, modulators, devices and systems have been developed in the context of optical communications. To illustrate \cut{the concept}\add{our combined approach featuring this trainable encoding,} we consider here a balanced-arm MZI with non-symmetric couplers, which is mathematically equivalent to nonlinear optical loop mirror (NOLM), and semiconductor optical amplifier (SOA) as nonlinear transformers of optical signal. In what follows, we use balanced-arm MZI with non-symmetric couplers and NOLM interchangeably.


\section{Methods} 
\subsection{Nonlinearity and effective dimensionality}
Many machine learning methods utilize non-linear mapping as part of the data manipulation process, for instance, radial basis functions, most variants of neural networks, kernel-based methods and boosting~\cite{Bishop2006}.  These mappings {\em can be made} in a higher-dimensional space, for instance, by having hidden layers that include more neurons than the input, but these are not carried out in a controlled and structured manner and are not explicitly used as part of the processing method.

At the heart of our method is the nonlinear mapping of input vectors to a high-dimensional space, which facilitates the application of various tasks. While our primary focus is on time series forecasting and prediction, which serves as an example of a regression task, we also demonstrate the versatility of our approach by applying it to a classification problem on a sub-sampled MNIST dataset. In spirit, our methods follow the rationale of SVM~\cite{cristianini2000introduction}, where both classification and regression tasks are made possible by mapping them to the high dimensional space where linear separation~\cite{Cover65} and approximate regression can be carried out; but, unlike SVM, where the nonlinear transformation is replaced by the corresponding kernel and support vectors should be identified, we rely on the speed of computing devices and their ability to carry out fast mapping and simple regression.

Clearly, the nature of the nonlinear mapping and its suitability for the data is a crucial but difficult aspect of the method that should be addressed. Relying on available optical telecommunication devices limits the nonlinear mapping we can utilize; nevertheless, one can apply different control parameters that govern the type of mapping achieved, as detailed in Sect.~\ref{sec:architecture}. Developing a principled approach for determining the optimal mapping parameters is beyond the scope of the current paper and will be the subject of future research.

\begin{figure*}[t]
\centering
\includegraphics[width=0.75\linewidth]{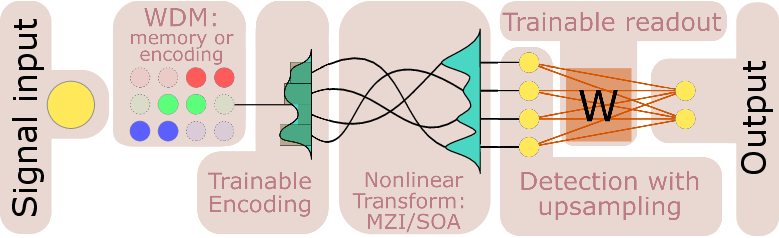}
\caption{Scheme of the proposed photonic ELMs/RCs.}
\label{fig: scheme}
\end{figure*}

\begin{figure*}[t]
\centering
\includegraphics[width=0.75\linewidth]{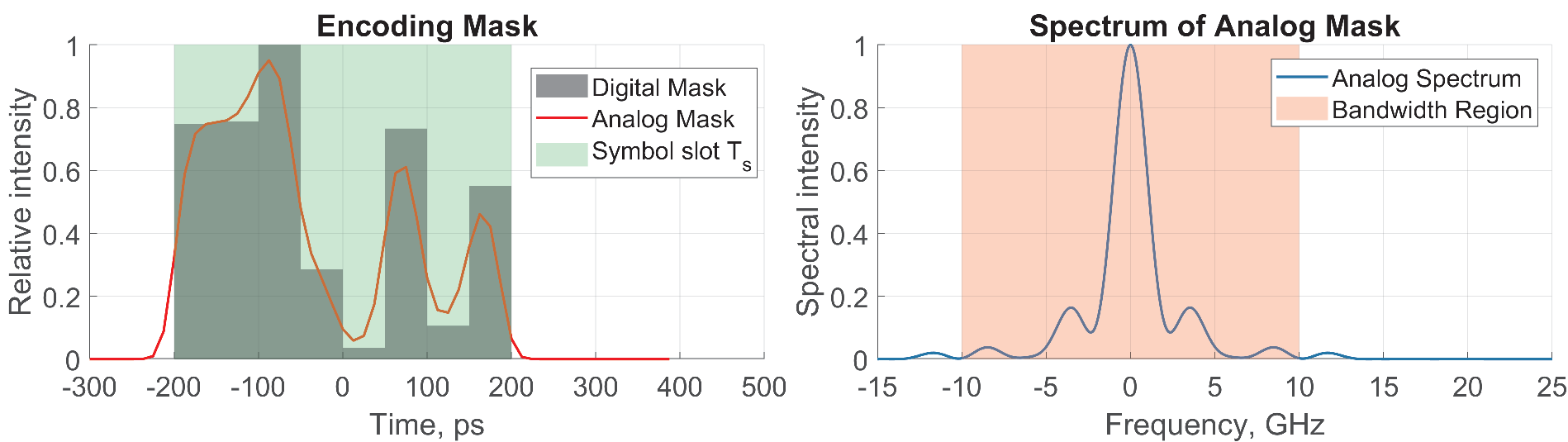}
\caption{Arbitrary and trainable encoding mask (digital and analog, limited by the bandwidth of arbitrary waveform generator). This example features eight trainable weights.}
\label{fig: encoding mask example}
\end{figure*}

For the more general case of noisy data and mapping process, it would be natural to assess the impact of the nonlinear mapping of inputs to the higher-dimensional space using entropic and mutual information measures. However, in this work, we use deterministic mapping and would like to ensure that the mapped data makes full use of the larger space and does not create trivial linear interdependencies. To do that, we will concentrate on the effective dimensionality of the mapped inputs. Transformation of the low-dimensional input signal into high-dimensional feature space will be suitable for computing \textit{only} if the output variables are linearly independent and are not redundant. The output signal should be spanned by nonlinear transformation across a large number of available dimensions to make them linearly separable.

While the mapping employed is complex and includes both nonlinearities and time-dependent components, one can employ linear algebra methodology to determine the effective dimensionality of the feature space. 
In statistical data analysis terms, the effective dimensionality of the mapped inputs is the number of orthogonal dimensions that would produce the same overall co-variation pattern.
This can be easily done using singular value analysis, exploratory factor analysis, principal component analysis and other dimensionality reduction techniques, both linear and non-linear~\cite{barberBRML2012}. One may also consider independent component analysis methods to explore the statistical independence property of mapped data. More suitable for measuring the complexity of time series are measures introduced in~\cite{Sigaki2018} and~\cite{Bandt2002}.
As the starting point for this research, we have adopted the singular value analysis, which is simple and effective. An alternative approach that could be explored is to enforce lower effective dimensionality through regularization during linear regression training of the weight matrix,
although this on its own cannot be used to evaluate the feature space dimensionality. In classification problems, dedicated algorithms to maximize the distance of data from the separating hyperplane could be employed~\cite{Anlauf_1989}.
\subsection{The architecture of the considered photonic ELM and RC systems}
\label{sec:architecture}
The considered photonic ELM/RC computing systems leverage the idea of using nonlinear transformation of signal into high-dimensional space for computational purposes. 
There are the following key steps in the computing architecture considered as schematically depicted in Figure \ref{fig: scheme}.  The first step is the interface between the real-world signal and an input into the computing system. Second, an input signal is tuned (modulated, coded) using available degrees of freedom of the system - parameters that can be used for tuning. There are two possibilities: In the case of slowly varying parameters, the signal undergoes a masking procedure to maximize the effective dimensionality of the system after the nonlinear transformation. When parameters are modulated 
fast characteristic of the proposed use of telecom devices), adjusting tuning to the incoming signal creates a possibility of dynamical, controllable masking. Third, the nonlinear element/system maps the input signal onto a high-dimensional feature space. Variables that cannot be linearly separated in a low-dimensional space can be successfully processed using linear algorithms in a high-dimensional space. The final step is signal processing in the high-dimensional space performed by the readout layer.
Below, we specify each of these steps considering the implementation of a general computing scheme using telecom-grade optical devices.

\subsubsection{Encoding techniques: dynamical masking}
To feed information into the proposed optical ELM/RC for computing one needs to encode the input data stream in an electromagnetic field. Consider that the input data has a form of a vector  $(s_1,s_2,...,s_N)$. Encoding of the input messages (symbols) onto the optical waveform can be done in different ways exploiting the available degrees of freedom of light. The advantage of the proposed ELM/RC is that we can use easily accessible and well-developed telecom devices. For example, amplitude and/or phase modulation can be employed to encode input symbols in the amplitude/phase of the optical signal in the time domain. Encoding can be done in different frequency channels using established WDM technology. In addition to the symbol encoding, the effective dimensionality of the input can be increased via additional signal-invariant modulation, i.e., masking. Consider a WDM-based encoding scheme with $L$ spectral channels and an optical pulse train having the form: 
\begin{equation}
    A(t) = \sqrt{P_0}\; \sum_{l=1}^{L}   \sum_{j=1}^{J} a_{l,j} \; g(t - j \cdot T_s) \exp\left(- i \omega_l t\right),
\label{eq: data stream to optics}
\end{equation}
where $P_0$ is the power scaling parameter, $L$ is a number of spectral channels (indexed by $l$) used for coding of the information, $T_s$ is a symbol rate, $a_{l,j}$ for $j=1,2,\dots$ is a symbol (in general, a complex number) in the spectral channel $l$ at the temporal position $j$  that is used for encoding the input information in the optical domain. It can be sampled from a discrete set (alphabet) or be continuous. 

Function $g(t)$ describes the shape of a carrier pulse (or an encoding mask) having the temporal scale $T_0$, which can be, in general, different from $T_s$. Thus, in the considered case, we need to encode the input data vector $(s_1,s_2,...,s_N)$ into array $a_{l,j}$. Evidently, this can be done in different ways, providing rich opportunities for the manipulation and optimization of the subsequent nonlinear mapping.For example, one can encode multiple consecutive symbols $s_i$ into multiple parallel instances of $a_{l,j}$, effectively parallelizing processing. Or one can introduce a time shift, when $a_{l,j}$ contain shifted copies of $s_i$ in different spectral channels, effectively introducing memory. To illustrate the general concept we consider here only simple intensity modulation ($a_{l,j}$ are real non-negative numbers).

\add{In this work, spectral multiplexing is used differently depending on the task. For time-series forecasting, we introduce memory across channels by encoding time-delayed replicas of a base symbol sequence: specifically, we define $a_{l,j} = a_{1,j-l+1}$ for $l = 1,\dots,L$ and $j \geq l$, such that channel $l$ contains the $(l-1)$-symbol delayed copy of the original sequence on channel 1. For classification tasks, we encode multiple features or input elements (image pixels) in parallel across different WDM channels at the same temporal index, effectively reducing the sequential processing length.}
\add{These encoding schemes are shown in Tab.~\ref{tab:wdm-encoding-forecasting-classification}}

\begin{table}[h]
\centering
\caption{WDM-based encoding schemes. (a) Time-series forecasting uses delayed copies across spectral channels to introduce memory. (b) Classification encodes features in parallel across channels to reduce sequential depth.}
\label{tab:wdm-encoding-forecasting-classification}

\vspace{0.5em}
\textbf{(a) Time-series forecasting}
\begin{tabular}{c|cccccc}
\toprule
WDM Channel & $j{=}1$ & $j{=}2$ & $j{=}3$ & $j{=}4$ & $j{=}5$ & $...$ \\
\midrule
$\omega_1$ & $s_1$ & $s_2$ & $s_3$ & $s_4$ & $s_5$ & ... \\
$\omega_2$ &       & $s_1$ & $s_2$ & $s_3$ & $s_4$ & ... \\
$\omega_3$ &       &       & $s_1$ & $s_2$ & $s_3$ & ... \\
$\omega_4$ &       &       &       & $s_1$ & $s_2$ & ... \\
$\omega_5$ &       &       &       &       & $s_1$ & ... \\
\bottomrule
\end{tabular}

\vspace{1.2em}
\textbf{(b) Classification}
\begin{tabular}{c|ccc}
\toprule
WDM Channel & $j{=}1$ & $j{=}2$ & $\cdots$ \\
\midrule
$\omega_1$ & $s_1$ & $s_6$ & $\cdots$ \\
$\omega_2$ & $s_2$ & $s_7$ & $\cdots$ \\
$\omega_3$ & $s_3$ & $s_8$ & $\cdots$ \\
$\omega_4$ & $s_4$ & $s_9$ & $\cdots$ \\
$\omega_5$ & $s_5$ & $s_{10}$ & $\cdots$ \\
\bottomrule
\end{tabular}
\end{table}

One can introduce asymmetry in the optical signal we use the skewed Gaussian pulse of the following form:
\begin{equation}
    g(t) = g(t,T_0,\alpha) = \frac{\exp(-\tau^2/2T_0^2)}{1 + \exp(-\alpha \tau/T_0)}
\end{equation}
where $\alpha$ is a skewness parameter, $\tau = (t - t_1) / t_2$ is the shifted and rescaled time introduced to align the mean and variance of the skewed function with those of a standard Gaussian (with $\alpha = 0$, $t_1 = 0$, $t_2 = 1$), scaling parameters $t_{1,2}$ adjust the shift and scale of the distribution, respectively. In this work, we always use $t$ parameters optimized to provide a skewed function that preserves the mean and variance of the original Gaussian distribution, thus allowing for a direct comparison while accounting for asymmetry in the mapping. The use of the asymmetric carrier pulse combined with up-sampling effectively plays a role in the masking procedure that assists the following nonlinear distribution of the same symbol into different parts of an output feature space.
~\\
Finally, the encoding mask can also be trained to optimize the performance of the ELM. In this case, we can tweak individual parameters of the encoding mask to maximize the accuracy of the proposed ELM. In this work, we used GWO \cite{mirjalili2014grey} for global optimization of the encoding mask and Nelder-Mead simplex method for refinement. Figure \ref{fig: encoding mask example} shows an example of an arbitrary encoding mask limited by the analog bandwidth of the arbitrary waveform generator.

\subsection{Nonlinear transformation in optical domain}

NOLM operation can be explained as follows: the input signal power is divided between the two arms of a waveguide loop (for instance, an optical fiber), the signal phase in each arm is changed by the nonlinear propagation, and the resulting signal is formed by coupling the output ports of the arms (see next subsection for details). NOLM can produce overall nonlinear response using unequal coupling ratios, creating asymmetry of the accumulated nonlinear phase shifts, as in the original proposal \cite{doran1988nonlinear} or by introducing imbalance in nonlinear propagation by using amplifiers (nonlinear amplifying loop mirror - NALM). Evidently, NOLM waveguide device can be realized on different material platforms. The same concept can be implemented in a nonlinear analog of the Mach–Zehnder interferometer, creating an interferometric phase converter to control the sign of the nonlinear phase shift \cite{Gabitov:02}. 

SOA is a well-developed technology with many attractive characteristics, including compact size, efficient electrical pumping, cost-effectiveness and wideband gain \cite{SOA01,SOA02,SOA03}. 
However, in high-speed optical communication applications, the nonlinear properties of SOAs, relatively slow gain recovery time, and comparatively high noise figures (compared to other optical amplifiers) pose serious challenges. 
The carrier dynamics of SOAs have a characteristic scale of several hundred picoseconds. In the context of high-speed optical communications, this produces dependence of an instantaneous SOA gain on the input optical signal power that results in patterning effects - nonlinear distortions with memory. However, these nonlinear and inherent memory features
can be attractive for optical computing applications, as demonstrated below.
We describe the transfer functions for both NOLM and SOA below. However, for clarity, we primarily present results based on NOLM-based computing. Examples of optical computing using SOA-based high-dimensional mapping can be found in \cite{Manuylovich:24}.  

The output state is obtained by taking the absolute value squared of the output of the complex transfer functions, i.e. we use intensity-only detection. We assume that the output signal is measured using photodetectors. Additionally, we consider the passive fiber losses are negligible compared to the signal modulation due to nonlinear signal transform.

\subsubsection {Nonlinear optical loop mirror}
The transfer function of a nonlinear optical loop mirror with a coupler having a split ratio $\kappa$ is given by the following expression \cite{doran1988nonlinear}:

\begin{equation}
\begin{aligned}
    A_{\text{out}}(t) 
    &= \sqrt{\kappa} \cdot \text{NLSE}\left(\sqrt{\kappa} A_{\text{in}}(t), \beta_2, \gamma, L_{\mathrm{NOLM}}\right) \\
    &+ i\sqrt{1-\kappa} \cdot \text{NLSE}\left(i\sqrt{1-\kappa} A_{\text{in}}(t), \beta_2, \gamma, L_{\mathrm{NOLM}}\right)
\end{aligned}
\label{eq:NOLM_transfer_function}
\end{equation}

Here the input light field $A_{\text{in}}(t)$ is split into two counter-propagating waves in the NOLM with the amplitudes defined by the coupling parameter $\kappa$ \cite{doran1988nonlinear}. 
The function $\text{NLSE}(A, \beta_2, \gamma, L_{\mathrm{NOLM}})$ here represents the solution of the nonlinear Schrödinger equation for the amplitude $A(t,z)$ with a given input $A_{in}(t)$ after propagating through a fiber of the length $L_{\mathrm{NOLM}}$, with a group velocity dispersion $\beta_2$ and nonlinearity parameter $\gamma$:

\begin{equation}
    \frac{\partial A}{\partial z} = -i\frac{\beta_2}{2} \frac{\partial^2 A}{\partial t^2} + i\gamma |A|^2 A
    \label{eq: NLSE}
\end{equation}

The intensity of the nonlinearly transformed signal can then be measured at the readout stage. 
When dispersive effects are negligible compared to nonlinear ones (e.g., using fiber/waveguide near zero-dispersion point or high power signal), then the transfer function is simplified to the compact form:

\begin{equation}
\nonumber
\begin{aligned}
    |A_{\text{out}}(t)|^2 &= |A_{\text{in}}(t)|^2 \left\{ 1 - 2 \kappa (1-\kappa) \,  \right. \\
&\quad \left. \times \left[ 1 + \cos \left( (1-2 \kappa) \gamma L_{\mathrm{NOLM}} |A_{\text{in}}(t)|^2 \right) \right] \right\}
\end{aligned}
\label{eq:NOLM_power_transfer_function}
\end{equation}

\begin{figure}[t]
\centering
\includegraphics[width=1\linewidth]{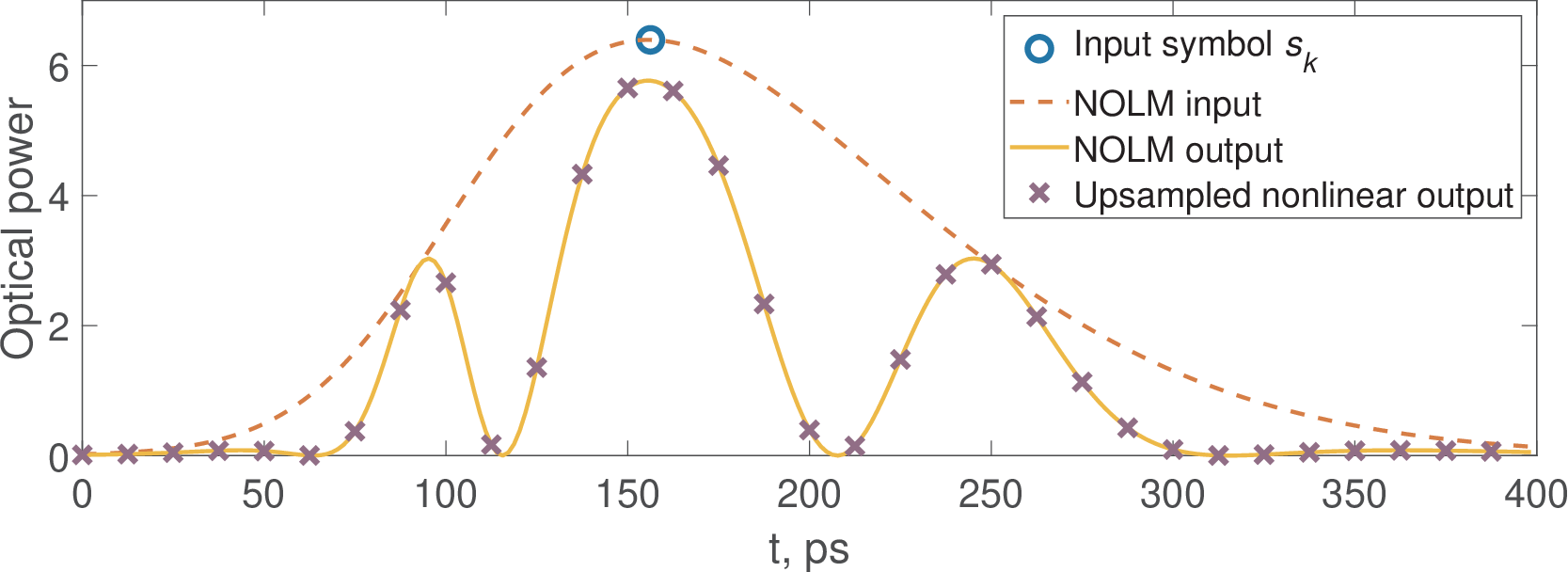}
\caption{Input symbols, their optical encoding, NOLM output, and the samples captured by a photodetector. \add{For readability, 16 samples per symbol are shown instead of the 32 used in experiments.}}
\label{fig:NOLM input-output}
\end{figure}

It is seen from this analytical approximation of the NOLM transfer function that by varying parameters $\kappa$ and $L_{\mathrm{NOLM}}$ one can dramatically change the properties of the nonlinear transformation.

In this work, to calculate the input-output signal transformation in nonlinear MZI, we numerically simulate signal propagation trough fiber by solving Eq.(...) using the Fourth-Order Runge–Kutta in the Interaction Picture Method~\cite{hult2007fourth}. We do it for both arms, i.e. with initial conditions $\sqrt{\kappa} A_{\text{in}}(t)$ and $i\sqrt{1-\kappa} A_{\text{in}}(t)$. We then combine the solutions at $z=L_{\mathrm{NOLM}}$ as in Eq.~(...) and take the absolute value squared to get the signal as registered by a photodetector.

\subsubsection {Semiconductor optical amplifier}
Nonlinear transformation of the input optical signal $A_{in}(t)= \sqrt{P_{in}(t)}\, \exp( i \phi_{in}(t))$ by SOA to the output field $A_{out}(t)= \sqrt{P_{out}(t)}\, \exp( i \phi_{out}(t))$ is governed by the well-established model \cite{SOA01,SOA02}:

\begin{equation}
\begin{aligned}
&P_{out}(t) = P_{in}(t) \exp[h(t)] \\
&\phi_{out}(t) = \phi_{in}(t)-\frac{\beta}{2} h(t)\\
&\frac{d h}{d t} = -\frac{h-h_0}{\tau_c}-\frac{P_{in}(t)}{E_{sat}}\left[\exp(h)-1\right]
\end{aligned}
\label{eq:SOA_ODEs}
\end{equation}

here in/out index denotes the input/output signal, $\beta$ is the linewidth enhancement (Henry) factor, $h_0$ parameter is related to the small signal gain $G_0=\exp(h_0)$, $\tau_c$ is the gain recovery time, $E_{sat}$ is a characteristic saturation energy. 
It is seen that the nonlinear transformation of the input temporal signal stream by SOA creates an effective device memory defined by the delayed gain recovery: gain at a certain point in time $h(t)$ depends on the signal in the previous moments. In our previous work~\cite{Manuylovich:24} we demonstrate that memory capacity associated with this effect to be $\geq3.5$. We demonstrate that this property can be exploited to create a high-dimensional feature output. Eqs.~\ref{eq:SOA_ODEs} are directly integrated using Runge-Kutta method.

By varying SOA current (linked to $h_0$ parameter), we can dynamically control the nonlinear transformation. 

\subsection{Readout approaches to implement high-dimensional feature space}
One of the key advantages of implementing nonlinear mapping to high-dimensional signals is the possibility to use simple processing, well-established telecom technologies and components at the readout layer. Though signal polarization and spatial modes can also be controlled and manipulated with telecom-grade devices, here we focus on the frequency and temporal domains at the output.

In the temporal domain, applying the up-sampling technique, data encoded in the symbol time interval (at the baud rate) and spread by the nonlinear transformation can be measured at the sampling rate of the receiver. When data encoding is implemented on both the amplitude and the optical phase of the carrier pulses, the standard telecom coherent receiver can be used to recover both amplitude and phase information at the sampling rate. Up-sampling here means the temporal sampling of the nonlinearly transformed signal (intensity in this considered illustration) at a higher rate than the encoding symbol (baud) rate. The up-sampling coefficient $M$ describes how many points we get for a single input pulse. Thus, the output of the considered ELM/RC is a nonlinearly transformed high-dimensional representation of the input signal, which is a key component of the computing. This process for NOLM and $M=16$ is illustrated in Fig.~\ref{fig:NOLM input-output}.

Applying a modulator at the sampling rate, we can adjust the readout weight in the optical domain or, using high-speed optoelectronics available in telecom, change weights in the electrical domain. Here, we simulate readout by multiplying the row-vector of intensities at the output of the ELM/RC with the regression column-vector, determined during the training procedure (see section~\ref{sec: training_validation}).

\cut{In the frequency domain, the separation of different spectral channels can be done using well-developed WDM technology. 
The output optical signal passes through a WDM de-multiplexer that separates different spectral channels in the optical domain. The temporal up-sampling readout approach is applied to each spectral channel in parallel.}

\cut{However, in this work, WDM is used only for encoding, and the channels are not separated after the nonlinear optical transform. Instead, we assume that all channels are detected together using a single photodetector at the output of the nonlinear photonic device.}

\add{In this work, we utilize WDM solely for encoding the input signal across multiple spectral channels. After the nonlinear transformation, these channels are not separated at the output stage. Instead, we assume a detection scheme where the combined optical output, containing all WDM channels, impinges on a single photodetector. The temporal up-sampling readout approach is then applied to this aggregated signal.}

\add{While this joint detection approach is employed here, it is worth noting that an alternative readout strategy \textit{could} involve separating the different spectral channels using well-developed WDM technology. In such a scenario, the output optical signal would pass through a WDM de-multiplexer, and the temporal up-sampling readout approach could be applied independently to each spectral channel in parallel. However, exploring such parallel processing via output WDM separation is beyond the scope of the current study, which focuses on the impact of WDM encoding combined with joint detection.}

\subsection{Training and validation} \label{sec: training_validation}

In this work, we show how the proposed approach can be utilized for the classical machine learning task of time series forecasting.
The general scheme of the proposed ELM/RC includes: (i) encoding of the input signal/vector onto the optical field, (ii) a nonlinear element that transforms the input signal (in this work we consider two examples:  the nonlinear loop mirror  \cite{doran1988nonlinear} and semiconductor optical amplifier),  (iii) trainable readout $W$ that includes detection with up-sampling. The trainable output layer $W$ is straightforward and easily implementable linear regression, enabling the device to be applied in time series forecasting. The scheme of the proposed device is shown in Fig. \ref{fig: scheme}.

We would like to stress that the proposed general concept is not limited to the particular choice of the element that implements nonlinear signal transformation. It can be implemented with a variety of nonlinear sub-systems. 

To train the system, we must provide examples of correct answers (targets) to given feature vectors. We pass multiple sections of the sequence to forecast through the system, collect the feature (row) vectors and assemble a so-called feature matrix $\mathbf{X}$. Then, we construct a (column) vector $\mathbf{Y}$ of correct answers or targets by putting the correct next symbol in front of the corresponding row of features. 

For example, if one wants to train the system to predict using $N$ symbols, one can use $M$ symbol sequences to construct the following feature matrix:
\begin{equation}
\begin{aligned}
s_1, s_2, \ldots, s_N & \rightarrow \begin{bmatrix} x_{11} & x_{12} & \cdots & x_{1K}  \end{bmatrix} \\
s_2, s_3, \ldots, s_{N+1} & \rightarrow  \begin{bmatrix} x_{21} & x_{22} & \cdots & x_{2K}  \end{bmatrix} \\
s_3, s_2, \ldots, s_{N+2} & \rightarrow  \begin{bmatrix} x_{31} & x_{32} & \cdots & x_{3K}  \end{bmatrix} \\
 & \quad \vdots \\
s_M, s_{M+1}, \ldots, s_{M+N} & \rightarrow \begin{bmatrix} x_{M1} & x_{M2} & \cdots & x_{MK} \end{bmatrix}
\end{aligned}
\end{equation}
Here, $K$ denotes the dimensionality of the output vector, $N$ denotes number of symbols fed into RC/ELM to generate features for next symbol prediction, $M$ denotes number of training samples and the symbol $\rightarrow$ denotes the procedure of encoding the symbols in a waveform, propagating this waveform through the physical system and detecting it. The row vectors on the right-hand side of the arrows are the feature vectors that correspond to the input symbol sequences.

When training to predict the next symbol, the corresponding target vector is:

\begin{equation}
\mathbf{Y}
\equiv
\begin{bmatrix} 
y_1 \\
y_2 \\
\vdots \\
y_M \\
\end{bmatrix}
=
\begin{bmatrix} 
s_{N+1} \\
s_{N+2} \\
\vdots \\
s_{M+N+1} \\
\end{bmatrix}
\end{equation}

In front of each (row) vector of features, we place the next symbol according to the sequence of symbols used to generate these features. This is what we refer to as the single-step prediction task. We also evaluated our approach for autoregressive multi-step prediction, where the single-step prediction is applied iteratively, updating the time frame to incorporate newly predicted symbols.

The process of training a model involves identifying the regression vector $\theta$ that minimizes the Mean Squared Error (MSE) between the predicted values, represented as $\mathbf{X}\theta$, and the actual target values, $\mathbf{Y}$. The dimensionality of $\theta$ corresponds to the number of output features, $K$. In the following, we also use the Normalized Mean Squared Error (NMSE), defined as the MSE divided by the variance of $\mathbf{Y}$. NMSE provides interpretability, as $1 - \text{NMSE} = R^2$, which represents the proportion of variance explained by our model. The goal is to find $\theta$ such that the difference between these predicted and actual values is as small as possible, which can be expressed as:

\begin{equation}
\min_\theta \| \mathbf{Y} - \mathbf{X} \theta \|^2_2
\end{equation}

To achieve this, an effective approach is to use the Moore-Penrose pseudoinverse, denoted by $\mathbf{X}^\dagger$. This pseudoinverse offers a least-squares solution to the equation $\mathbf{Y} = \mathbf{X} \theta$. When the matrix $\mathbf{X}$ is decomposed using Singular Value Decomposition (SVD) as $\mathbf{X} = \mathbf{U} \mathbf{\Sigma} \mathbf{V}^*$, the pseudoinverse can be computed as:

\begin{equation}
\mathbf{X}^\dagger = \mathbf{V} \mathbf{\Sigma}^{-1} \mathbf{U}^*
\end{equation}

In this context, $\mathbf{\Sigma}^{-1}$ is formed by taking the reciprocal of each nonzero singular value in $\mathbf{\Sigma}$ and transposing the resulting matrix. Using this pseudo-inverse, the optimal $\theta$ is given by:

\begin{equation}
\theta = \mathbf{X}^\dagger \mathbf{Y}
\end{equation}

However, calculating the pseudo-inverse involves inverting the singular values, which can amplify even minor variations in the feature matrix, leading to significant fluctuations in the regression vector. This issue is particularly problematic when noise is present in the data, as it can result in unstable and unreliable predictions. Regularization techniques are employed to mitigate this problem.

We use a widely used regularization technique, $L_2$ regularization, which penalizes large coefficients by adding a term proportional to the square of the coefficient magnitudes. This can be implemented by truncating the singular values in the SVD of $\mathbf{X}$, which reduces the influence of smaller singular values associated with less important features.

In truncated SVD, a threshold is set, and singular values below this threshold are zeroed out, effectively lowering the rank of $\mathbf{\Sigma}$. This approach can be represented as solving the following regularized problem:

\begin{equation}
\min_\theta \| \mathbf{Y} - \mathbf{X} \theta \|^2_2 + \lambda_{\text{reg}} \| \theta \|^2_2
\end{equation}

Here, $\lambda_{\text{reg}}$ is a regularization parameter that balances the trade-off between fitting the data well and keeping $\theta$ small. The adjusted singular values, denoted by $\mathbf{\Sigma}_r$, are defined as:

\begin{equation}
\mathbf{\Sigma}_r = \text{diag}(\sigma_1, \sigma_2, \ldots, \sigma_r, 0, \ldots, 0)
\end{equation}

where $r$ is the number of singular values retained. This leads to a regularized estimate of $\theta$:

\begin{equation}
\theta_r = \mathbf{V}_r \mathbf{\Sigma}_r^{-1} \mathbf{U}_r^* \mathbf{Y}
\end{equation}

In this expression, $\mathbf{V}_r$ and $\mathbf{U}_r$ correspond to the matrices associated with the first $r$ singular values.

We carefully adjusted $\lambda_{\text{reg}}$ to minimize MSE on the validation subset.

In what follows, we train a simple linear regression model on this high-dimensional output vector space. To achieve good performance of the proposed computing system, the high-dimensional output vector space must be non-degenerate, providing distinct and linearly independent vectors for diverse inputs. The feature matrix in the regression consists of output vectors, where the number of features, or components of these vectors, determines the dimensionality. However, this perceived dimensionality can be misleading due to the potential degeneracy among the columns, reducing the effective dimensionality of the output. 
When some columns are linear combinations of others, they do not contribute new information, are useless for regression, and lead to a low-rank feature matrix. This shows the importance of characterizing the true dimensionality of the space represented by the matrix, which can significantly differ from just the number of columns. We use singular value decomposition to analyze the effective dimensionality as it provides insight into the matrix structure by factorizing it into two unitary matrices and a diagonal one containing singular values. These singular values, forming the singular value spectrum, indicate the significance of each dimension in capturing the matrix's variability.

To evaluate and demonstrate the depth of the feature space formed by the optical encoding and nonlinear transformation, we introduce 1000 input symbols $x_k$ from 0 to 1 through the PELM and construct a feature matrix using output row vectors. 

\begin{figure}[t]
\centering
\includegraphics[width=1\linewidth]{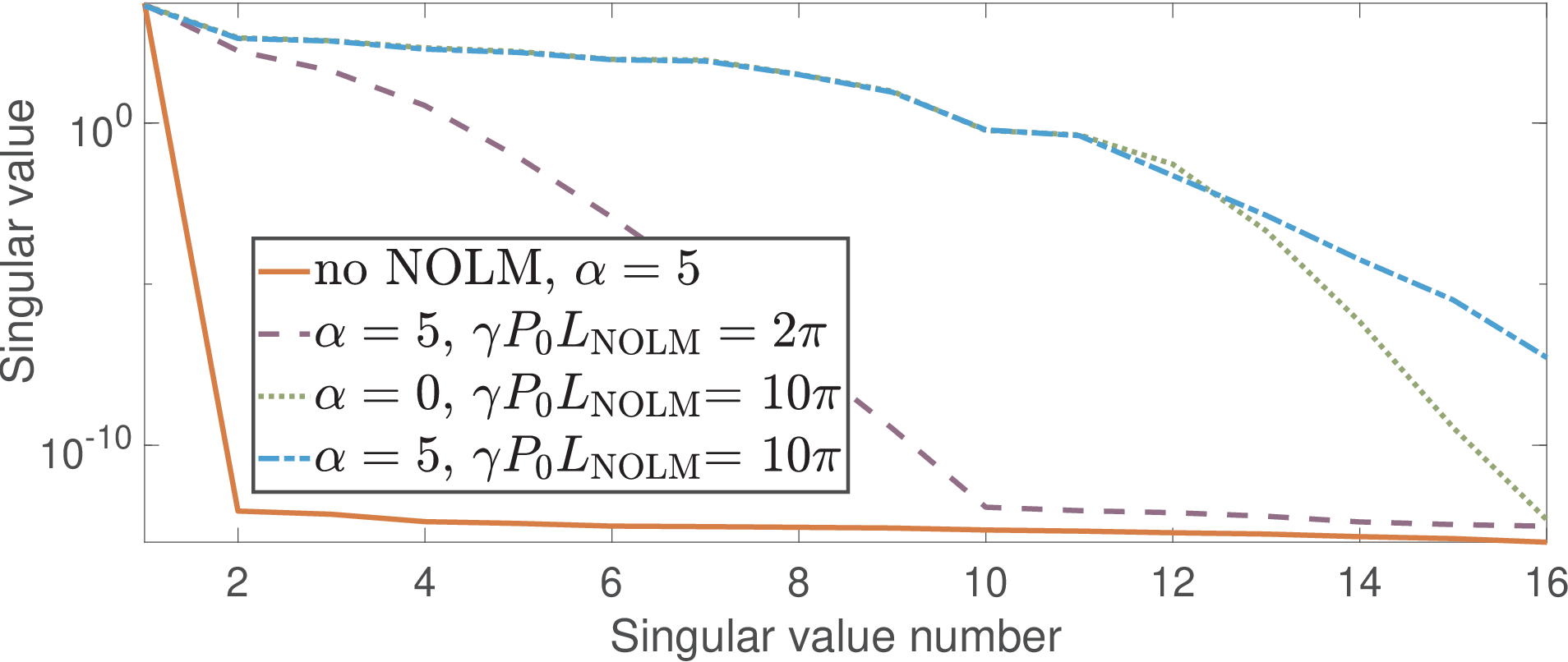}
\caption{Improvement in the spectrum of singular values of the feature matrix after applying NOLM. The graph shows a richer singular value spectrum at NOLM output, indicating higher effective dimensionality.}
\label{fig: singular value spectra}
\end{figure}

We illustrate the dependence of the singular value spectrum of the feature matrix on $\gamma P_0 L_{\mathrm{NOLM}}$ and $\alpha$ in Fig. \ref{fig: singular value spectra}, and also present a feature matrix derived from up-sampled signals that were not processed through the NOLM. One can see that the feature space for simply encoded and up-sampled input symbols is highly degenerate, as all feature variability can be explained with a single singular value: the second singular value is more than 10 orders of magnitude lower than the first (see no NOLM line in Fig.~\ref{fig: singular value spectra}). This corresponds to the effective rank 1 feature space. When NOLM is used for nonlinearly dispersing the signal, the (non-)degeneracy of the feature space depends on the nonlinear phase shift and symmetry of the encoding function $g$. This comparison reveals how the NOLM significantly influences the inherent dimensionality of the output state.
When applying the WDM technique for the readout procedure, one can retrieve a full 2D map of features in time-frequency axes.

\subsection{Simulation hyperparameters}
To ensure reproducibility and clarity, we summarize below all hyperparameters used across our numerical experiments in Table~\ref{tab:hyperparams}. These include system-specific parameters for the encoding and nonlinear transformation stages, as well as settings related to the learning process and readout. The table covers both nonlinear MZI/NOLM- and SOA-based implementations, along with details of the receiver and training configuration.
\\
\renewcommand{\arraystretch}{1.2}
\setlength{\tabcolsep}{5pt}
\begin{table}[!h]
\centering
\footnotesize
\begin{tabular}{>{\raggedright}p{1.7cm} >{\raggedright}p{3.4cm} l}
\toprule
\textbf{Category} & \textbf{Parameter} & \textbf{Value} \\
\midrule 
Encoding & Symbol interval $T_s$& \cut{80}\add{400} ps \\
         & Samples per Symbol & 32 \\
         & Peak power $P_0$ & 50 W \\
         & Num WDM channels & 5 \\
         & \add{WDM}\cut{F}requency step $\Delta\nu$ & 50 GHz \\
         & Encoding mask size & 8 \\
         & Analog bandwidth & 20 GHz \\
MZI/NOLM & Nonlinearity $\gamma$ & $0.8 \; \text{W}^{-1}\text{km}^{-1}$  \\
         & GVD $\beta_2$ & $26$ $\text{ps}^2\text{km}^{-1}$ \\
         & Length L & 100 m \\
         & Coupling ratio $\kappa$ & 0.3 \\
SOA      & Recovery time $\tau$ & 200 ps \\
         & Henry factor $\beta$ & 5 \\
         & Log gain $h_0$ & 6.91 \\
         & Saturation energy $E_\text{sat}$ & 8 pJ \\         
Receiver & Sampling rate & 80 GSa/s \\
         & Samples per symbol & 32 \\
Learning & Training symbols & 5000 \\
         & Testing symbols & 100-5000 \\
         & Symbols $N$ & 15 \\
         & \add{Regularization $\lambda_{\text{reg}}$ (optimized per task and device SOA/NOLM)} &  \add{$10^{-8}$ to $10^{-2}$} \\
         & \cut{Regularization $\lambda_{\text{reg}}$ }& \cut{0.01} \\

\bottomrule
\end{tabular}
\caption{All hyperparameters used in the simulation.}
\label{tab:hyperparams}
\end{table}


\section{Benchmarks}
For evaluating the performance, we used three different time series with different levels of complexity and the MNIST classification benchmark. When simulating the series, we employ typical parameters commonly used for the evaluation of machine learning models, specifically using the Mackey-Glass series in the form:
\begin{equation}
    \frac{dx}{dt} =  \frac{0.2x(t - 17)}{1 + x(t - 17)^{10}} - 0.1 x(t)
\label{eq: Mackey-Glass equation}
\end{equation}

And another time series is given by the solution of the Rossler attractor:
\begin{equation}
\begin{aligned}
    dx/dt &= -y - z, \\
    dy/dt &= x + 0.2y, \\
    dz/dt &= 0.2 + z(x - 5.7)
\end{aligned}
\label{eq: Rossler attractor equation}
\end{equation}


These dynamical systems are known for being chaotic and are often used for estimating the performance of forecasting algorithms \cite{chomiak2024time}. For 3D attractor, we used only the $x$ component for training and testing the prediction accuracy. The inherent unpredictability of chaotic systems, characterized by their sensitivity to initial conditions, makes them ideal benchmarks for testing the limits of predictive models. In such environments, even the slightest variation in initial conditions can lead to vastly different outcomes, challenging the algorithms to capture the complex dynamics at play.

To solve Eq.~\ref{eq: Mackey-Glass equation}, we used the $\texttt{dde23}$ MATLAB solver with an adaptive step size to ensure target relative and absolute tolerances of $\texttt{RelTol=1e-6}$ and $\texttt{AbsTol=1e-8}$. The solution was then interpolated onto a uniform grid with a step size of 1, following the approach in Jaeger and Haas \cite{jaeger2004harnessing}. For the R\"ossler attractor, we used the $\texttt{ode45}$ MATLAB solver with adaptive step to ensure the same relative and absolute tolerances. We then interpolated the solution onto a uniform grid with a step size of 0.25. For the R\"ossler attractor, the time series was rescaled to lie within the range $[0.1, 1.1]$ to avoid zero or negative values, which are incompatible with optical power levels in our simulation. In contrast, the Mackey–Glass time series was used without normalization, as its amplitude remained within a suitable range for our modeling.

\begin{figure}[t]
\centering
\includegraphics[width=1\linewidth]{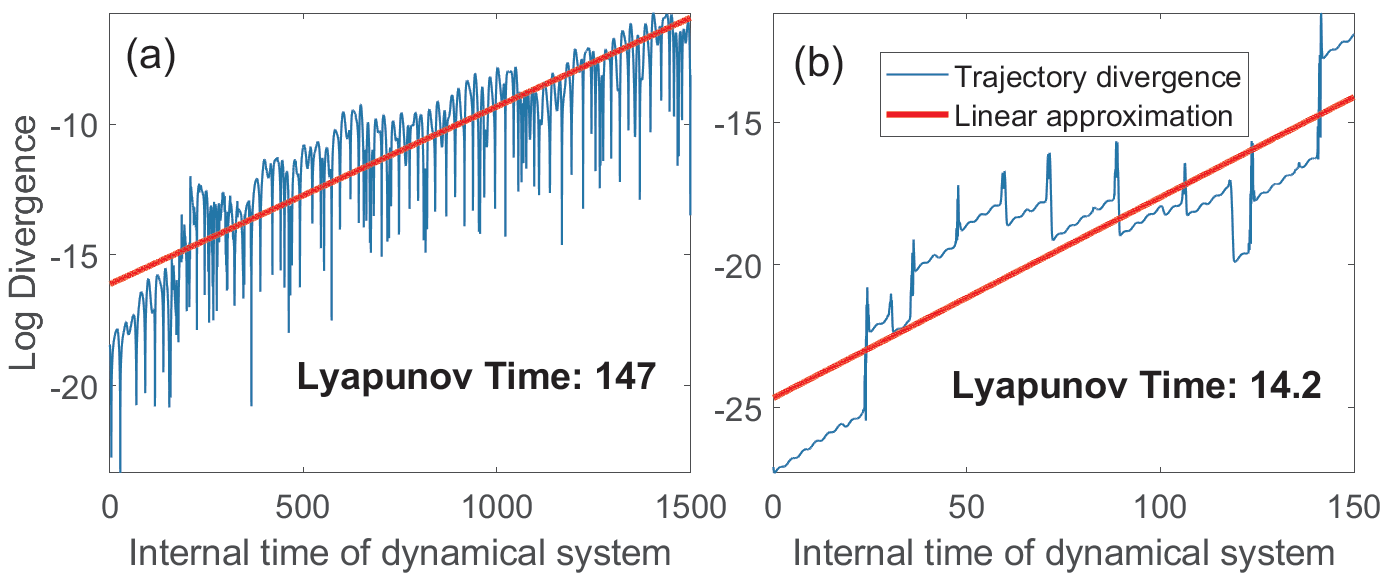}
\caption{Comparison of Lyapunov times for different dynamical systems, illustrating how increased chaoticity, from the (a) Mackey-Glass series to (b) Rossler attractor, results in reduced Lyapunov time and increased unpredictability.}
\label{fig: Lyapunov times}
\end{figure}

To characterize the randomness of these systems, we estimate the Lyapunov time, which measures the rate at which nearby trajectories in the system's phase space diverge. Specifically, the Lyapunov time is inversely related to the Lyapunov exponent of the system, indicating how quickly initial uncertainties or errors grow over time. To estimate the Lyapunov time for each dynamical system, we numerically integrated them from slightly different initial conditions and calculated the $\textsc{L}_2$ norm of the difference of these trajectories over time. These trajectories for the considered systems are shown in Fig. \ref{fig: Lyapunov times}. The Lyapunov time was estimated as the inverse slope of the difference in the logarithmic scale. In what follows, we show how the increased chaoticity of and reduction in Lyapunov time leads to reduced prediction ability of the proposed PELM. Chaoticity increases from the Mackey-Glass series to the R\"ossler attractor, with a corresponding decrease in Lyapunov time, indicating faster unpredictability in systems like R\"ossler due to exponential growth of initial differences, as shown in Fig. \ref{fig: Lyapunov times}.

To validate our Lyapunov time estimation based on trajectory divergence, we also computed the Lyapunov exponent for the R\"ossler attractor using the standard variational method based on linearization of the system equations. Specifically, we integrated both the original system and its linearized form (variational equations) and periodically orthonormalized the perturbation vectors using QR decomposition to estimate the largest Lyapunov exponent. This yielded a value for inverse Lyapunov exponent of approximately 13.4, which is consistent with the rate inferred from the divergence of nearby trajectories shown in Fig.~\ref{fig: Lyapunov times}. For the Mackey–Glass system, a similar analysis could not be performed due to its infinite-dimensional nature as a delay differential equation, which complicates the formulation and integration of variational equations. Therefore, for Mackey–Glass, we relied solely on the trajectory-based method to estimate the Lyapunov time.

For the classification task,  we took 10,000 samples from the MNIST dataset and split them 50/50 for training/testing. Since linear classifiers already achieve around 90\% accuracy on the original 784-pixel images, we increased the task difficulty by downsampling the images to just 15 pixels. We accomplished this using column-pivoting QR decomposition to identify and retain the most informative pixels. To do this, we reshaped the MNIST dataset into a matrix in the form 10,000 by 784 and applied column-pivoting QR decomposition, following the data-driven QR sensing paradigm, which is basically a compressed sensing approach performed on a tailored basis~\cite{manohar2018data}. The pivoting QR factorization seeks to reorder the columns such that the most "informative" columns are moved to the forefront. The measure of "informativeness" here is rooted in the magnitudes of the diagonal entries of the upper triangular matrix $\mathbf{R}$ resulting from the QR factorization. Columns, or, in our case, pixels, corresponding to larger diagonal entries in $\mathbf{R}$ are deemed more significant. Thus, we select the most informative columns. To ensure a proper evaluation, this decomposition was performed exclusively on the training subset to prevent any information from the testing subset from influencing the pre-processing stage. The identified pixel positions were subsequently used to downsample images in the testing subset. We used Linear Discriminant Analysis (LDA) for multi-class classification using either original selected pixels (linear classification) or processed by the photonic ELM. Applying this approach, we observed an accuracy of 42\% on the testing subset using linear classification. The 15 selected pixels were then input into the ELM to achieve a high-dimensional nonlinear mapping. A linear classifier was trained on the mapped samples from the training subset within this higher-dimensional feature space. The performance of the classifier was then evaluated using the mapped samples from the testing subset. When using WDM for encoding, we simultaneously transmit 5 different pixels across 5 separate spectral channels, effectively sending the input in batches of 5 pixels at a time, with 3 such batches per image.

\section{Results and discussion}

To test the ability of the approach to capture the dynamics of the considered systems, we used regularized linear regression over the feature space of the output of the PELM. Regularization parameter was optimized for long-term autoregressive prediction by testing the long-term prediction accuracy of the model for regularization parameters from $10^{-10}$ to $10^{-2}$, and choosing the value that provides the best accuracy on the autoregressive prediction. So the value of the regularization parameter was selected to avoid overfitting the local prediction while making the model capable of predicting the long-scale dynamics of the three test dynamical systems. Regularized pseudo-inversion of the feature matrix was used to learn the regression vector from the training data: regularization is performed by replacing all singular values that are below the current regularization parameter with zeros. Singular values and pseudoinversion are calculated by using singular value decomposition of the feature matrix. Then, we segment the validation data into sequences that the model will use to make predictions. Each sequence includes a set of input symbols followed by the symbols to be predicted. The number of sequences is determined based on the size of the validation data and the length of the sequences to be predicted. Typically, we used 10-15 validation sequences. For each input sequence, we predict the sequence of symbols that follow by taking an input sequence and, in a loop, making a prediction for the next symbol in the sequence.
After each prediction, the input sequence is updated by removing its first symbol and appending the predicted one. This updated sequence is then used for the next iteration.
This process is repeated for the length of the pre-defined autoregressive prediction length. We used different autoregressive prediction lengths for different dynamical systems: from 1000 for the Mackey-Glass system to 250 for the R\"ossler attractor.

The simulation parameters that we used for training and testing the proposed system, are given below. We set the receiver sampling rate of 80 GSa/s, with 32 samples recorded per symbol to provide the up-sampling of the encoded signal.

We used symbol-to-pulse train encoding with Gaussian, skewed Gaussian, or a trainable mask. The input pulse had a peak power of 50 W. Once more, we would like to stress that this power corresponds to NOLM based on SMF and can be substantially scaled down by using highly nonlinear fiber, or integrated MZI/NOLM with other material platform. When employing the trainable mask, it was first set digitally with ideal (infinitely steep) transitions, then simulated for encoding by an electro-optical modulator with a 20 GHz analog bandwidth. For WDM multi-channel spectral encoding, we used a frequency step of $\Delta\nu=50~\text{GHz}$ between channels, with 5 channels corresponding to frequency shifts of 0, $-\Delta\nu$, $+\Delta\nu$, $-2\Delta\nu$, and $+2\Delta\nu$. For time series forecasting, we encoded time-shifted copies of the same symbol sequences in other spectral channels so that channel 2 contains a 1-symbol delayed copy of the channel 1 sequence, channel 3 contains a 2-symbol delayed copy of the channel 1 sequence, etc. A similar approach was used for memory enhancement in RC in microresonators \cite{castro2024wavelength}. However, unlike the approach in \cite{castro2024wavelength}, our scheme additionally employs trainable encoding. Theoretical studies \cite{hulser2022role} have also explored memory enhancement via input-stage modifications, particularly in SOAs and microring resonators. Our work extends these insights by demonstrating an implementation in a WDM-based framework, offering flexible spectral encoding, trainable masking for dynamic input modulation, and a scalable approach using commercially available telecom components. We did not demultiplex the WDM channels at the output and instead simulated the readout as being measured directly from the nonlinear transform’s output using a single photodiode. Using WDM for the classification task, we simply encoded different pixels into different WDM channels in parallel, thus reducing the number of sequential symbols to be fed into the ELM.

The NOLM was characterized using parameters of a single mode fiber: a nonlinear coefficient $\gamma=0.8 \times 10^{-3} ~\mathrm{W^{-1} \, m^{-1}}$, group velocity dispersion coefficient $\beta_2=26 \times 10^{-3} ~\mathrm{ps^2 \, m^{-1}}$, fiber length $L_{\mathrm{NOLM}}=100 ~\text{m}$, and the coupling ratio $\kappa= 0.3$. Evidently, employing highly nonlinear fibers or other material platforms, all resulting optical pulse parameters can be easily scaled.

For the time series forecasting task, the model was trained using 4,000 symbols and tested on 100-1000 symbols depending on the chaoticity of the time series. We used batches of $N=15$ symbols to predict the next one. To make a multi-step prediction, we utilized the autoregressive approach, when the newly predicted symbols are used to predict the next ones. A regularization amplitude was optimized for performance and to prevent overfitting. For the classification task, the model was trained using 5,000 images from the training set, which were sub-sampled using column-pivoting QR decomposition. The pixel locations chosen for the training subset were then applied to the testing subset during the testing phase, which also included 5,000 sample images.

\begin{figure}[t]
\centering
\includegraphics[width=1\linewidth]{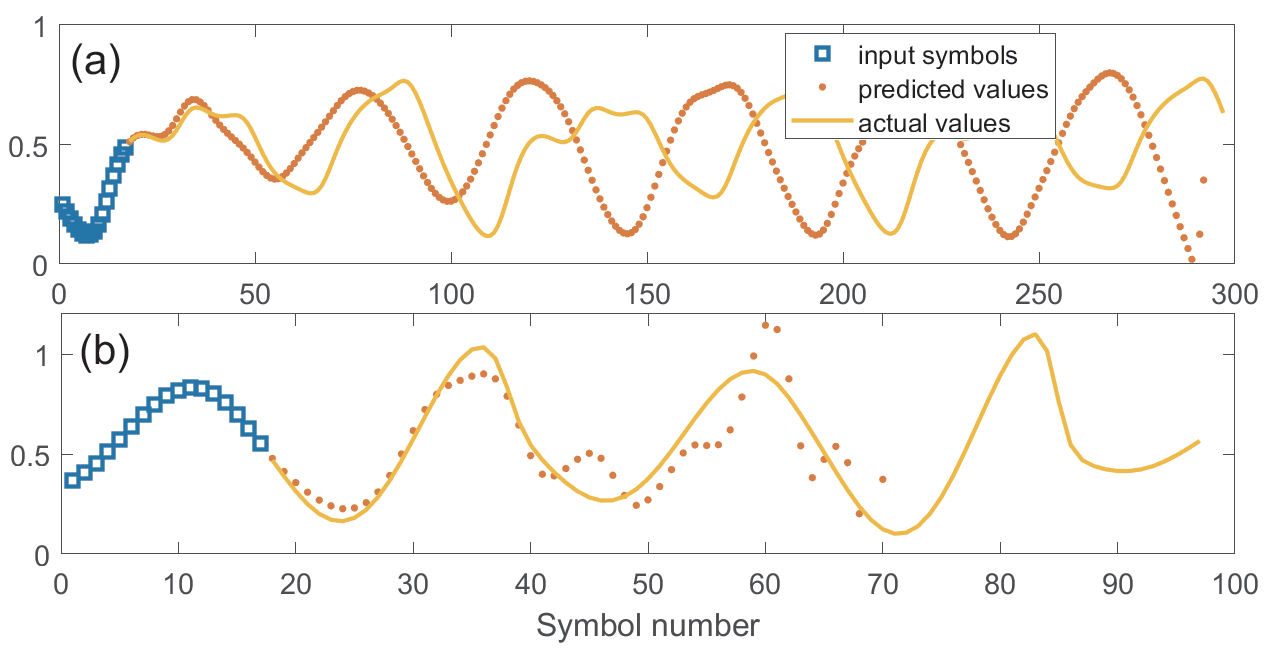}
\caption{Performance, for comparison, of the linear model on time-series prediction tasks. Poor prediction accuracy is achieved for all the Mackey-Glass (a) and R\"ossler (b) time series, which indicates the inability of the linear model to capture the complex dynamics of these systems.}
\label{fig:linear_time_series_prediction}
\end{figure}

We began by applying simple linear regression and linear classification to the upsampled raw data to establish a reference performance baseline. This initial approach served as a benchmark for evaluating the improvements offered by the nonlinear systems used. As expected, the linear regression model exhibited poor performance across both chaotic time-series prediction tasks and the sub-sampled MNIST classification task. Specifically, the tasks included autoregressive predictions on the Mackey-Glass (MG) and Rössler (R) time series, with results illustrated in Fig.~\ref{fig:linear_time_series_prediction}.

Although the single-step prediction NMSE appeared relatively low, at $\text{NMSE}_{\text{MG}} \approx 5 \cdot 10^{-6}$, $\text{NMSE}_{\text{R}} \approx 2 \cdot 10^{-4}$, the model failed at long-term prediction. This failure was due to the linear system's inability to capture and replicate the underlying dynamics of these complex systems.

\begin{figure}[!h]
\centering
\includegraphics[width=1\linewidth]{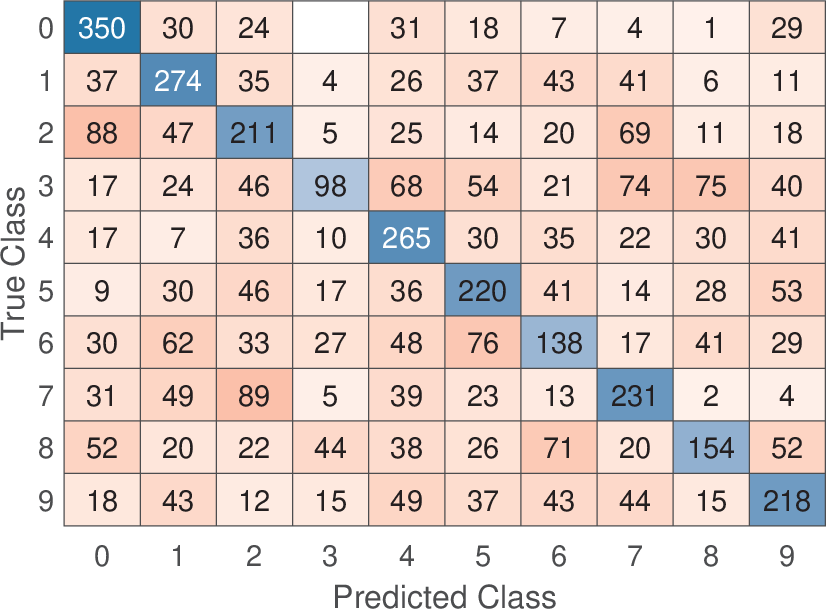}
\caption{Confusion matrix for the linear model on the sub-sampled MNIST classification task. The high number of off-diagonal entries indicates frequent misclassification.}
\label{fig:confusion_matrix_linear_classification}
\end{figure}

For the classification task on the sub-sampled MNIST dataset, the linear model's performance was similarly limited. The model struggled to separate the digit classes, which led to high misclassification rates. This shortfall is evident when examining the confusion matrix in Fig.~\ref{fig:confusion_matrix_linear_classification}, where off-diagonal elements are prominent, showing a high misclassification rate.

The overall accuracy on the test subset for the sub-sampled MNIST is only $42\%$.

\begin{figure}[!h]
\centering
\includegraphics[width=1\linewidth]{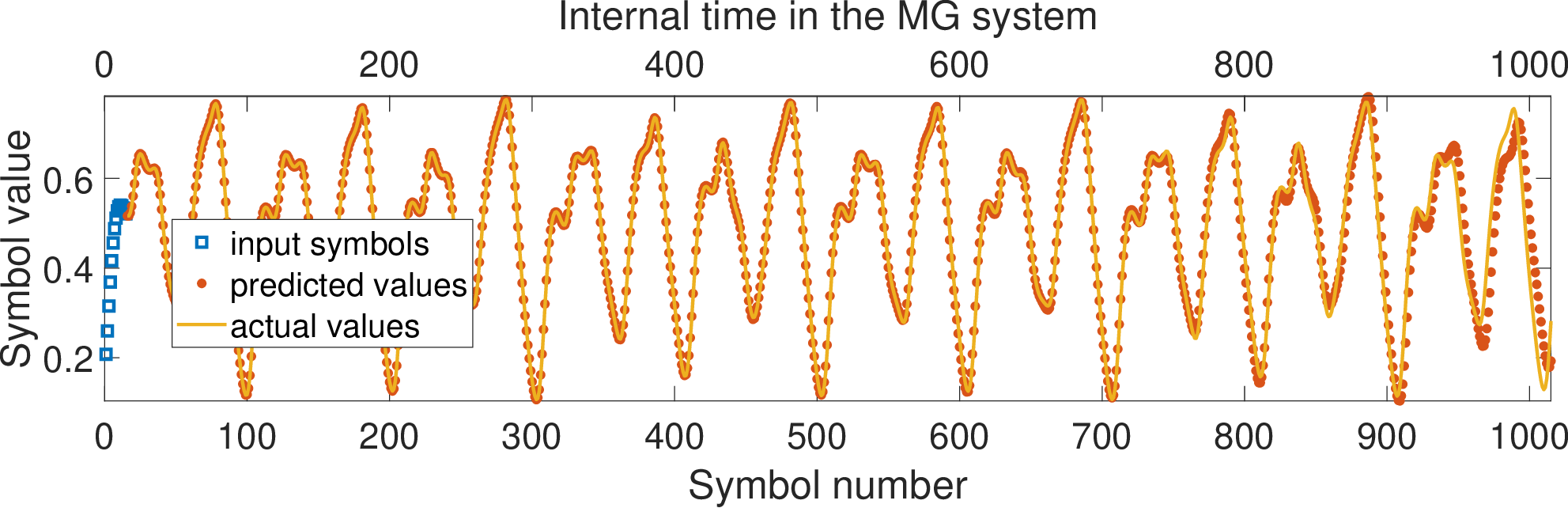}
\caption{Example of autoregressive prediction of Mackey-Glass series for 1000 symbols using PELM.}
\label{fig:MG_example_with_PELM}
\end{figure}

Now we determined the baseline accuracy of the linear models on the chosen tasks, and we can compare the accuracy achieved using PELM with the parameters described above. Figure~\ref{fig:MG_example_with_PELM} shows an example of the increased predictive ability of the model when high-dimensional nonlinear mapping is enabled via passing the encoded signal through the PELM.

NMSE averaged over multiple predicted sequences, similar to shown in Fig.~\ref{fig:MG_example_with_PELM} but taken from different parts of the Mackey-Glass sequence, is equal to $\text{NMSE}=0.097$ for the autoregressive prediction depth of 1000 symbols, which is a significant improvement compared to the purely linear model, shown in Fig.~\ref{fig:linear_time_series_prediction}(a).

\begin{figure}[!h]
\centering
\includegraphics[width=1\linewidth]{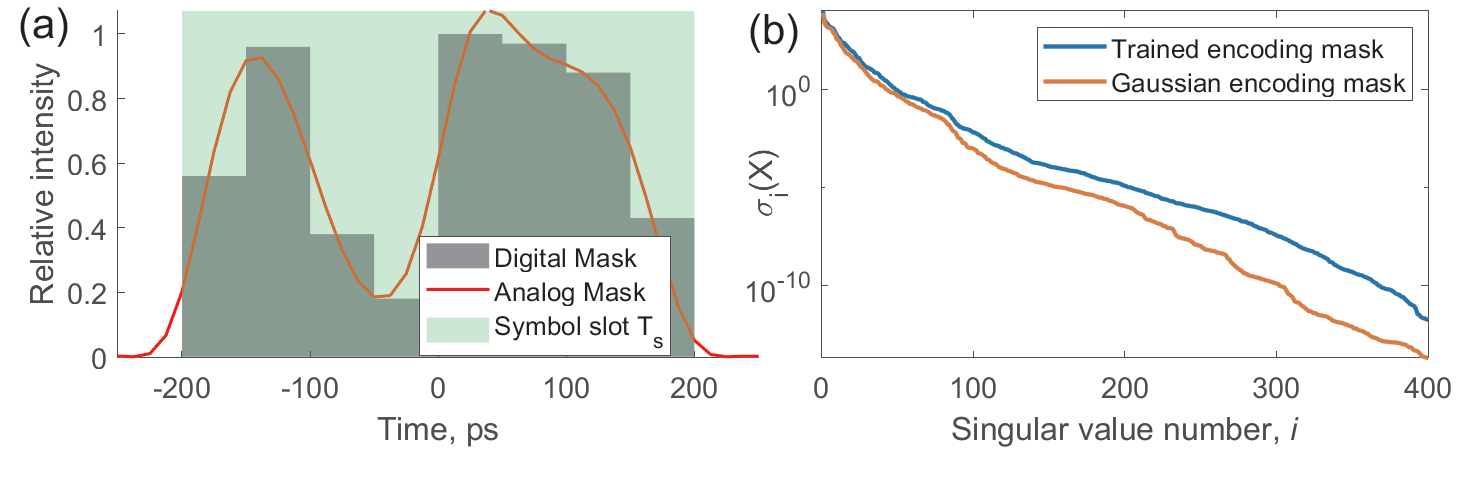}
\caption{Trained encoding mask (a) and corresponding singular spectrum change (b) in the feature matrix of the PELM model. The model performance was enhanced by using an 8-slot trainable encoding mask, allowing each symbol to be encoded with 8 independently optimized amplitudes.}
\label{fig:trained_mask_and_svd_combined}
\end{figure}

The performance of the PELM can be improved further by employing a trainable encoding mask instead of using a simple Gaussian shape for the optical encoding of the input symbols. We employed an 8-slot trainable encoding mask, such that each symbol is encoded with 8 independent amplitudes. The waveform of each corresponding pulse is then described as a convolution of this arbitrary-shaped vector of length 8 with a finite-bandwidth response function of the modulator. We used the raised cosine function as the response function, and the analog bandwidth of the modulator was set to 20 GHz. The encoding mask was trained by a combination of global and local optimization techniques. We used GWO \cite{mirjalili2014grey} for global optimization and Nelder-Mead simplex method for local optimization. The trained mask and the corresponding change in the singular spectrum of the feature matrix are shown in Fig.~\ref{fig:trained_mask_and_svd_combined}. The corresponding improvement in performance for autoregressive prediction depth of 1000 symbols is reduced to $\text{NMSE}=0.036$, a 2.7 times improvement compared to simple Gaussian encoding mask.

When using another telecom technology, WDM, for encoding (i.e., combining multiple signals within a single optical channel before passing them through the PELM), performance can also be improved compared to the simpler single-channel encoding approach. When using 5 channels with 50 GHz separation and employing single-symbol shift, as described above, we managed to achieve $\text{NMSE}=0.054$ for simple Gaussian encoding and $\text{NMSE}=0.024$ for trained encoding mask. The encoding mask used was the same for all spectral channels. The performance can be further optimized by using individual trainable encoding masks for each WDM channel, but this requires further research and training via higher-order parameter optimization, which is beyond the scope of this work. 

\begin{figure}[!h]
\centering
\includegraphics[width=1\linewidth]{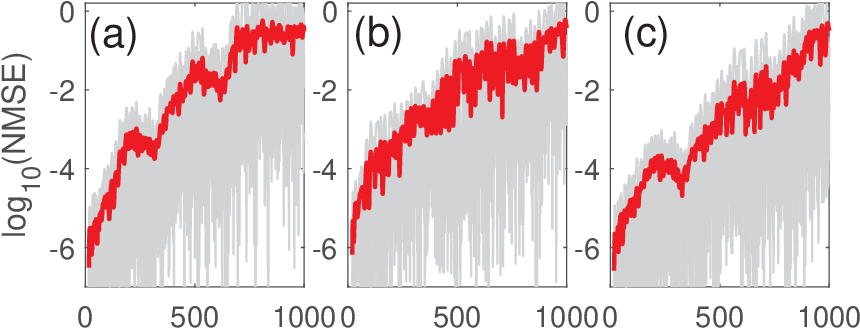}
\caption{NMSE as a function of autoregressive prediction depth for different encoding techniques: (a) single-channel Gaussian pulse encoding, (b) 5-channel WDM with Gaussian pulse encoding, and (c) single-channel trained encoding mask. Errors for individual predicted sequences are shown in gray, and the average error across all predicted sequences is highlighted in red.}
\label{fig:MG_errors_all_methods}
\end{figure}

Figure~\ref{fig:MG_errors_all_methods} shows how NMSE depends on autoregressive prediction depths for different encoding techniques: single channel with Gaussian pulse encoding (a), 5-channel WDM with Gaussian pulse encoding (b), and single channel with trained encoding mask (c). The errors for individual predicted sequences are depicted in gray, while the average error across all predicted sequences is highlighted in red.

Compared to other photonic RC approaches for Mackey–Glass prediction, our method outperforms \cite{dong2019optical}, achieving an NMSE of 0.01 for predicting 300 symbols (six quasi-periods), whereas \cite{dong2019optical} reports an NMSE of 0.1. In comparison to the delay-based RC in \cite{ortin2015unified}, our approach achieves similar accuracy for single-step prediction; however, a direct comparison of multi-step prediction performance is not possible due to the lack of reported error metrics in \cite{ortin2015unified}.

\begin{figure}[!h]
\centering
\includegraphics[width=1\linewidth]{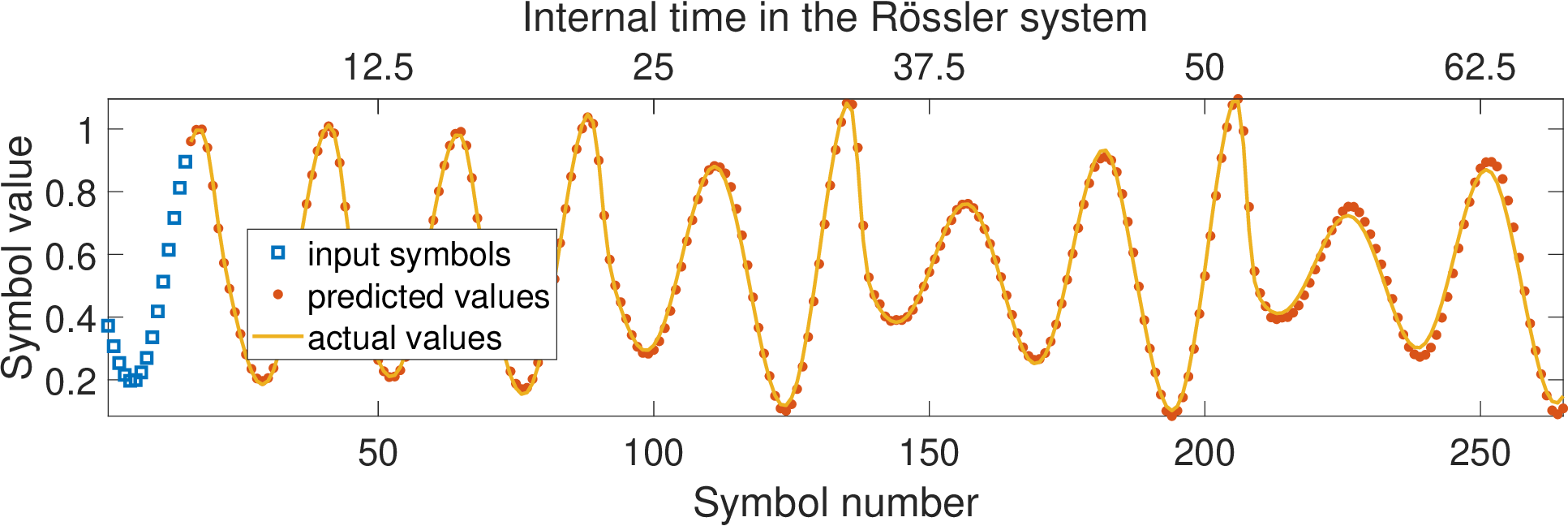}
\caption{Example of autoregressive forecasting of the R\"ossler time series using a single-channel Gaussian-shaped encoding.}
\label{fig:Rossler_example}
\end{figure}

When testing on the R\"ossler attractor data, we similarly can get boost in prediction accuracy when utilizing multiple WDM channels and trainable encoding mask. Fig.~\ref{fig:Rossler_example} shows an example of autoregressive forecasting of the Rossler time series using single channel Gaussian-shape encoding.

\begin{figure}[!h]
\centering
\includegraphics[width=1\linewidth]{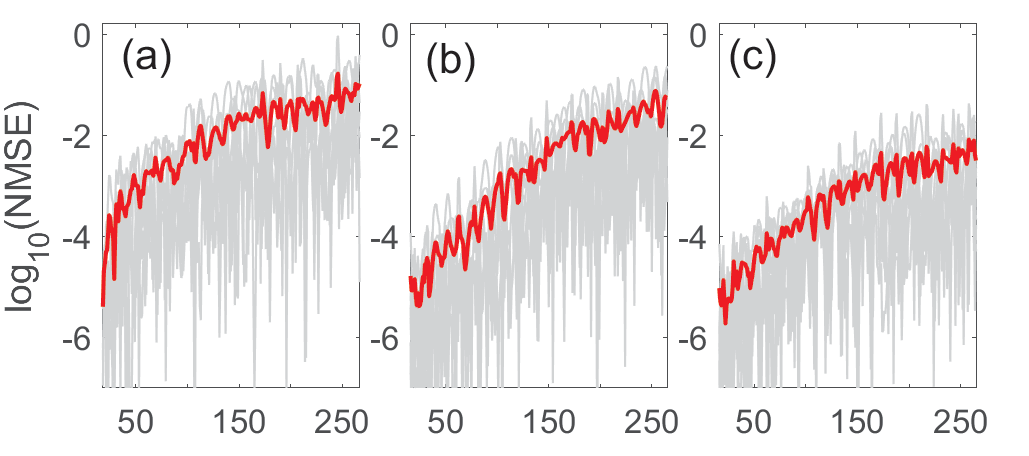}
\caption{NMSE versus autoregressive prediction depth for the R\"ossler time series (a); the results demonstrate reduced prediction error when using 3 WDM channels (b) or a trainable encoding mask (c), compared to single-channel encoding.}
\label{fig:Rossler_errors_all_methods}
\end{figure}

Using 3 WDM channels for encoding or a trainable encoding mask reduces NMSE across autoregressive prediction depths, as shown in Fig.~\ref{fig:Rossler_errors_all_methods}.

The overall accuracy for R\"ossler time series prediction using NOLM again can be improved from $\text{NMSE}_R=1.16\times10^{-1}$ to $\text{NMSE}_R=5.7\times10^{-2}$ when using WMD encoding and to $\text{NMSE}_R=9.79\times10^{-3}$ when using trained encoding mask even with a single encoding channel.

\begin{figure}[!h]
\centering
\includegraphics[width=1\linewidth]{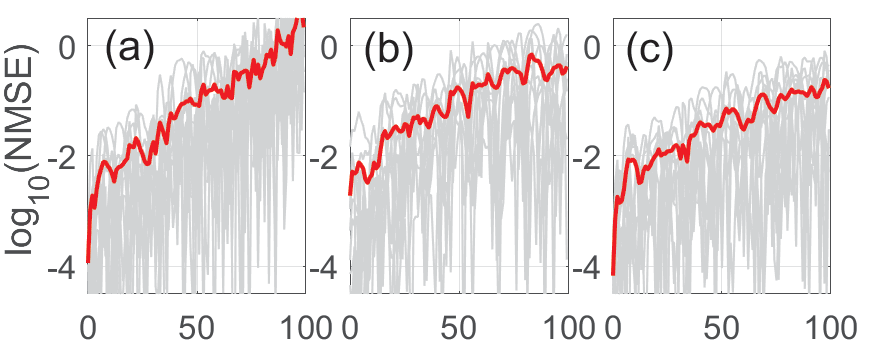}
\caption{NMSE versus autoregressive prediction depth when using SOA as a nonlinear mapping device, for the R\"ossler time series (a); the results demonstrate reduced prediction error when using 3 WDM channels (b) or a trainable encoding mask (c), compared to single-channel encoding.}
\label{fig:Rossler_SOA_errors_all_methods}
\end{figure}

We also tested the proposed approach on SOA as a nonlinear mapping device. All system parameters are kept the same except the peak power $P_0=0.7~\text{W}$. SOA parameters used for simulation (see eq.~\ref{eq:SOA_ODEs}) are the following: $\beta=5$, small signal gain $30$ dB, $\tau_c = 200$ ps, $E_{sat}=8$ pJ. All time series and computing parameters are identical to those in Fig.~\ref{fig:Rossler_errors_all_methods}, except that the nonlinear transform is now performed using an SOA. Figure~\ref{fig:Rossler_SOA_errors_all_methods} shows how applying WDM encoding (b) and trainable encoding mask (c) outperform the standard Gaussian encoding mask (a) for R\"ossler attractor prediction task.

\begin{figure}[!h]
\centering
\includegraphics[width=1\linewidth]{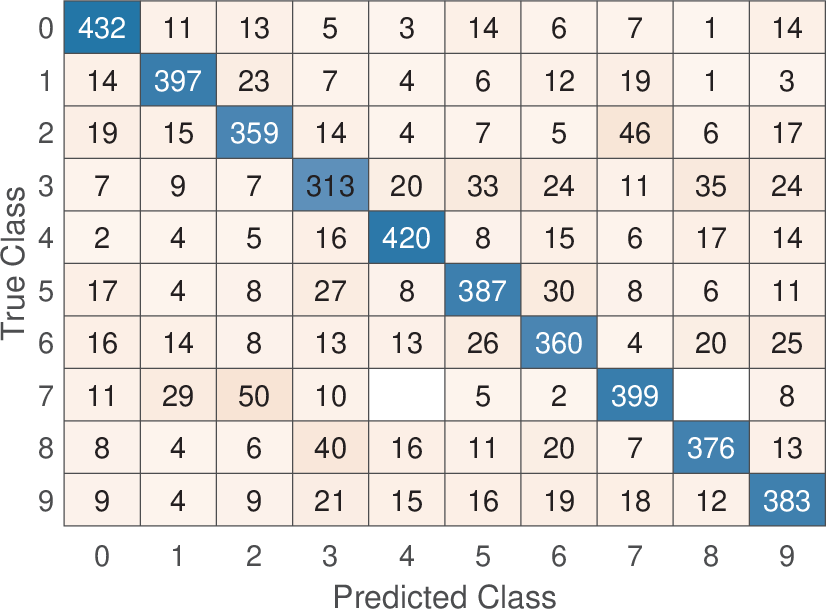}
\caption{Confusion matrix for the NOLM-based Extreme Learning Machine (ELM) model on the sub-sampled MNIST classification task. The model shows significantly improved performance compared to the linear model, with a reduced number of off-diagonal misclassifications.}
\label{fig:confusion_matrix_ELM_classification}
\end{figure}

Finally, In the classification task, the proposed approach (see Fig.~\ref{fig:confusion_matrix_ELM_classification} for the improved confusion matrix) demonstrates a substantial performance improvement over the baseline linear model on the sub-sampled MNIST classification task. With WDM encoding, we achieve comparable performance, with the added advantage of encoding multiple symbols simultaneously across different spectral channels.

Testing on the sub-sampled MNIST classification task shows an overall test accuracy of $77\%$ for the NOLM-based ELM model, a significant increase from the linear model's $42\%$ accuracy.


\section{Conclusion}
We proposed and demonstrated through numerical modeling photonic ELM and RC designs based on the well-developed telecommunication technology and component platform. The important advantage of this approach is the combination of low-cost, high-speed characteristics of linear and nonlinear elements with the frequency parallelism techniques well-established in wave-division-multiplexing optical communication systems. High-capacity and high-speed optical communication are utilized to create a large number of degrees of freedom in the time-frequency domain for controllable encoding and relatively low-power parallel nonlinear mapping of the input signal into high-dimensionality output for linear processing. The use of multiple frequency channels for encoding input information is important for the efficiency of nonlinear mapping. Indeed, in linear time-invariant optical systems, the frequency components in the output and input signals are the same. The nonlinear optical transformation leads to frequency component mixing and generation of new harmonics providing conditions for efficient mapping to higher-dimensional space. 

It is important to point out that while we focused here on the forecasting of time series and a classical classification task, a similar projection to the high-dimensional case is also highly relevant and can be employed for various other tasks following the methodology of SVM~\cite{cristianini2000introduction}. The difference, however, is that the proposed approach does not require the use of kernel-based methods. Key to improving the success of the non-linear projection used is adjusting the mapping parameters in a way that will provide an easier separation of features. This is an aspect of the research that is still ongoing. 
\bibliographystyle{ieeetr}
\bibliography{references}

\end{document}